\documentclass[aps,prd,twocolumn,preprintnumbers,
superscriptaddress,showpacs,floatfix]{revtex4}

\usepackage{bm}
\usepackage{epsfig}
\usepackage{graphics}
\usepackage{amsmath}

\begin{document}

\preprint{UT-STPD-11/01}

\def\beq{\begin{equation}}
\def\eeq{\end{equation}}
\def\bea{\begin{eqnarray}}
\def\eea{\end{eqnarray}}

\newcommand{\Eref}[1]{Eq.~(\ref{#1})}
\newcommand{\Sref}[1]{Sec.~\ref{#1}}
\newcommand{\Fref}[1]{Fig.~\ref{#1}}
\newcommand{\Tref}[1]{Table~\ref{#1}}
\newcommand{\cref}[1]{Ref.~\cite{#1}}
\newcommand{\Erefs}[2]{Eqs.~(\ref{#1}) and (\ref{#2})}
\newcommand{\beqs}{\begin{subequations}}
\newcommand{\eeqs}{\end{subequations}}

\def\lf{\left(}
\def\rg{\right)}
\newcommand{\etal}{{\it et al.\/}}
\newcommand{\GeV}{{\mbox{\rm GeV}}}
\newcommand{\pb}{{\mbox{\rm pb}}}
\def\mcr{{\tt micrOMEGAs}}

\newcommand{\bmm}{{\ensuremath{{\rm BR}\lf B_s\to \mu^+\mu^-\rg}}}
\newcommand{\bsg}{{\ensuremath{{\rm BR}\lf b\to s\gamma\rg}}}
\newcommand{\btn}{{\ensuremath{{\rm R}\lf B_u\to \tau\nu\rg}}}
\newcommand{\Dam}{{\ensuremath{\delta a_{\mu}}}}
\newcommand{\Omx}{{\ensuremath{\Omega_{\rm LSP} h^2}}}
\newcommand{\ssi}{{\ensuremath{\sigma^{\rm SI}_{\tilde\chi p}}}}
\newcommand{\ssd}{{\ensuremath{\sigma^{\rm SD}_{\tilde\chi p}}}}
\newcommand{\mx}{{\ensuremath{m_{\rm LSP}}}}
\newcommand{\Dst}{{\ensuremath{\Delta_{\tilde\tau_2}}}}
\newcommand{\spn}{{\ensuremath{\sigma_{\pi N}}}}
\newcommand{\mst}{{\ensuremath{m_{\tilde\tau_2}}}}
\newcommand{\Mg}{{\ensuremath{M_{1/2}}}}
\newcommand{\AMg}{{\ensuremath{A_0/M_{1/2}}}}

\title{CMSSM with Yukawa Quasi-Unification Revisited}

\author{N. Karagiannakis}
\email{nikar@auth.gr} \affiliation{Physics Division, School of
Technology, Aristotle University of Thessaloniki, Thessaloniki
54124, Greece}
\author{G. Lazarides}
\email{lazaride@eng.auth.gr} \affiliation{Physics Division, School
of Technology, Aristotle University of Thessaloniki, Thessaloniki
54124, Greece}
\author{C. Pallis}
\email{kpallis@auth.gr} \affiliation{Department of Physics,
University of Cyprus, P.O. Box 20537, CY-1678 Nicosia, CYPRUS}

\date{\today}

\begin{abstract}
The constrained minimal supersymmetric standard model with $\mu>0$
supplemented by an `asymptotic' Yukawa coupling quasi-unification
condition, which allows an acceptable $b$-quark mass, is
reinvestigated. Imposing updated constraints from the cold dark
matter abundance in the universe, $B$ physics, the muon anomalous
magnetic moment, and the mass $m_h$ of the lightest neutral
CP-even Higgs boson, we find that the allowed parameter space is
quite limited but not unnaturally small with the cold dark matter
abundance suppressed only via neutralino-stau coannihilations. The
lightest neutralino with mass in the range $(341-677)~\GeV$ is
possibly detectable in the future direct cold dark matter searches
via its spin-independent cross section with nucleon. In the
allowed parameter space of the model, we obtain
$m_h=(117-122.2)~\GeV$.

\end{abstract}

\maketitle

\section{Introduction}\label{sec:intro}

The well-known \emph{constrained minimal supersymmetric standard
model} (CMSSM) \cite{Cmssm0, Cmssm,cmssm1,cmssm2}, which is a
highly predictive version of the \emph{minimal supersymmetric
standard model} (MSSM) based on universal boundary conditions for
the soft \emph{supersymmetry} (SUSY) breaking parameters, can be
further restricted by being embedded in a SUSY \emph{grand unified
theory} (GUT) with a gauge group containing $SU(4)_c$ and
$SU(2)_R$. This can lead \cite{pana} to `asymptotic' \emph{Yukawa
unification} (YU) \cite{als}, i.e. the exact unification of the
third generation Yukawa coupling constants $h_t$, $h_b$, and
$h_\tau$ of the top quark, the bottom quark, and the tau lepton,
respectively, at the SUSY GUT scale $M_{\rm GUT}$. The simplest
GUT gauge group which contains both $SU(4)_c$ and $SU(2)_R$ is the
\emph{Pati-Salam} (PS) group $G_{\rm PS}=SU(4)_c\times
SU(2)_L\times SU(2)_R$ \cite{leontaris,jean} -- for YU within
$SO(10)$, see Refs. \cite{baery,raby}.

However, given the experimental values of the top-quark and
tau-lepton masses (which, combined with YU, naturally restrict
$\tan\beta\sim50$), the CMSSM supplemented by the assumption of YU
yields unacceptable values of the $b$-quark mass $m_b$ for both signs
of the parameter $\mu$. This is due to the presence of sizable
SUSY corrections \cite{copw} to $m_b$ (about 20$\%$), which arise
\cite{copw,pierce} from sbottom-gluino (mainly) and stop-chargino
loops and have the sign of $\mu$ -- with the standard sign
convention of Ref.~\cite{sugra}. The predicted tree-level
$m_b(M_Z)$, which turns out to be close to the upper edge of its
$95\%$ \emph{confidence level} (c.l.) experimental range receives,
for $\mu>0$ [$\mu<0$], large positive [negative] corrections which
drive it well above [a little below] the allowed range.
Consequently, for both signs of $\mu$, YU leads to an unacceptable
$m_b(M_Z)$ with the $\mu<0$ case being much less disfavored.

The usual strategy to resolve this discrepancy is the introduction
of several kinds of nonuniversalities in the scalar \cite{baery,
raby} and/or gaugino \cite{nath,shafi} sector of MSSM with an
approximate preservation of YU. On the contrary, in
Ref.~\cite{qcdm}, concrete SUSY GUT models based on the PS gauge
group are constructed which naturally yield a moderate deviation
from exact YU and, thus, can allow acceptable values of the
$b$-quark mass for both signs of $\mu$ within the CMSSM. In
particular, the Higgs sector of the simplest PS model
\cite{leontaris, jean} is extended so that the electroweak Higgs
fields are not exclusively contained in a $SU(2)_L\times SU(2)_R$
bidoublet superfield but
receive subdominant contributions from other representations too.
As a consequence, a moderate violation of YU is naturally obtained,
which can allow an acceptable $b$-quark mass even with universal
boundary conditions. It is also remarkable that the resulting
extended SUSY PS models support new successful versions
\cite{axilleas} of hybrid inflation based solely on renormalizable
superpotential terms.

These models provide us with a set of `asymptotic' Yukawa
quasi-unification conditions which replace exact YU. However,
applying one of these conditions in the $\mu<0$ case does not
lead \cite{muneg,nova} to a viable scheme. This is due to the
fact that the parameter space allowed by the \emph{cold dark
matter} (CDM) requirements turns out \cite{muneg, nova} to lie
lower than the one allowed by the inclusive decay
$b\rightarrow s\gamma$ in the $m_{\rm LSP}-\Delta_{\tilde\tau_2}$
plane, where $m_{\rm LSP}$ is the mass of the lightest sparticle
(LSP), which, in our case, is the lightest neutralino $\tilde\chi$
and
$\Delta_{\tilde\tau_2}=(m_{\tilde\tau_2}-m_{\rm LSP})/m_{\rm LSP}$
is the relative mass splitting between the LSP and the lightest
stau mass eigenstate $\tilde\tau_2$. This result is
strengthened by the fact that $\mu<0$ is strongly disfavored by
the constraint arising from the deviation $\delta a_\mu$ of the
measured value of the muon anomalous magnetic moment $a_\mu$
from its predicted value $a^{\rm SM}_\mu$ in the
\emph{standard model} (SM). Indeed, $\mu<0$ is defended only
at 3$\sigma$ by
the calculation of $a^{\rm SM}_\mu$ based on the $\tau$-decay
data, whereas there is a stronger and stronger
tendency at present to prefer the $e^+e^-$-annihilation data for
the calculation of $a^{\rm SM}_\mu$, which favor the $\mu>0$
regime. Given the above situation, we focus here on the $\mu>0$
case.

Let us recall that, in this case, the suitable `asymptotic'
Yukawa quasi-unification condition applied \cite{qcdm, nova}
is
\begin{equation}
h_t:h_b:h_\tau=|1+c|:|1-c|:|1+3c|. \label{minimal}
\end{equation}
This relation depends on a single parameter $c$, which is
taken, for simplicity, to be real and lying in the range $0<c<1$.
With fixed masses for the fermions of the third generation, we
can determine the parameters $c$ and $\tan\beta$ so that
Eq.~(\ref{minimal}) is satisfied. In contrast to the original
version of the CMSSM \cite{Cmssm, cmssm1, cmssm2}, therefore,
$\tan\beta$ is not a free parameter but it can be restricted,
within our set-up, via \Eref{minimal} to relatively large values.
The remaining free parameters of our model are the universal soft
SUSY breaking parameters defined at $M_{\rm GUT}$, i.e.,
\begin{equation}
M_{1/2},~~m_0,~~\mbox{and}~~A_0, \label{param}
\end{equation}
where the symbols above denote the common gaugino mass, scalar
mass, and trilinear scalar coupling constant, respectively.
These parameters can be restricted by employing a number of
experimental and cosmological requirements as in Refs.~\cite{qcdm,
nova}. In view of the expected data from the \emph{Large Hadron
Collider} (LHC), it would be worth to retest our model
against observations using the most up-to-date version of the
available constraints.

We exhibit the cosmological and phenomenological
requirements which we considered in our investigation in
Sec.~\ref{sec:pheno} and we restrict the parameter space of our
model in Sec.~\ref{results}. Finally, we test our model from the
perspective of the CDM direct detection experiments in
Sec.~\ref{det} and summarize our conclusions in Sec.~\ref{con}.

\section{Cosmological and Phenomenological Constraints}
\label{sec:pheno}

In our investigation, we integrate the two-loop renormalization
group equations for the gauge and Yukawa coupling constants and
the one-loop ones for the soft SUSY breaking parameters between
$M_{\rm GUT}$ and a common SUSY threshold $M_{\rm SUSY}
\simeq(m_{\tilde t_1}m_{\tilde t_2})^{1/2}$ ($\tilde t_{1,2}$ are
the stop mass
eigenstates) determined in consistency with the SUSY spectrum. At
$M_{\rm SUSY}$, we impose radiative electroweak symmetry breaking,
evaluate the SUSY spectrum employing the publicly available
calculator {\tt SOFTSUSY} \cite{Softsusy}, and incorporate the SUSY
corrections to the $b$ and $\tau$ mass \cite{pierce}. The
corrections to the $\tau$-lepton mass $m_\tau$ (almost 4$\%$) lead
\cite{qcdm, muneg} to
a small decrease of $\tan\beta$. From $M_{\rm SUSY}$ to $M_Z$, the
running of gauge and Yukawa coupling constants is continued using
the SM renormalization group equations.

The parameter space of our model can be restricted by using a
number of phenomenological and cosmological constraints.  We
calculate them using the latest version of the publicly available
code {\tt micrOMEGAs} \cite{micro}. We now briefly discuss these
requirements -- for similar recent analyses, see Ref.~\cite{lhc}
for CMSSM or Refs.~\cite{baerlhc, shafi} for MSSM with YU.

\paragraph{\hspace*{-0.3cm} SM Fermion Masses.} The masses of the
fermions of the third
generation play a crucial role in the determination of the
evolution of the Yukawa coupling constants. For the $b$-quark
mass, we adopt as an input parameter in our analysis the
$\overline{\rm MS}$ $b$-quark mass, which at 1$\sigma$ is
\cite{pdata}
\beq m_b \lf m_b\rg^{\overline{\rm MS}}=4.19^{+0.18}_{-0.06}~\GeV.
\eeq
This range is evolved up to $M_Z$ using the central value
$\alpha_s(M_Z)=0.1184$ \cite{pdata} of the strong fine structure
constant at $M_Z$ and then converted to the ${\rm \overline{DR}}$
scheme in accordance with the analysis of Ref.~\cite{baermb}. We
obtain, at $95\%$ c.l.,
\beq 2.745\lesssim  m_b(M_Z)/{\rm GeV}\lesssim 3.13
\label{mbrg}\eeq
with the central value being $m_b(M_Z)=2.84~\GeV$. For the
top-quark mass, we use the central pole mass ($M_t$) as an input
parameter \cite{mtmt}:
\beq M_t=173~\GeV~~\Rightarrow~~m_t(m_t)=164.6~\GeV\eeq
with $m_t(m_t)$ being the running mass of the $t$ quark. We also
take the central value $m_{\tau}(M_Z) = 1.748~\GeV$ \cite{baermb}
of the ${\overline{\rm DR}}$ tau-lepton mass at $M_Z$.

\paragraph{\hspace*{-0.3cm} Cold Dark Matter Considerations.}
\label{phenoa}
According to the WMAP results \cite{wmap}, the $95\%$ c.l. range
for the CDM abundance is
\beq \Omega_{\rm CDM}h^2=0.1126\pm0.0072. \label{cdmba}\eeq
In the context of the CMSSM, the LSP can be the lightest
neutralino $\tilde\chi$ and naturally arises as a CDM
candidate. We require its relic abundance $\Omega_{\rm LSP}h^2$
in the universe
not to exceed the upper bound derived from Eq.~(\ref{cdmba}) --
the lower bound is not considered since other production
mechanisms \cite{scn} of LSPs may be present too and/or other
CDM candidates \cite{axino, Baerax} may also contribute to
$\Omega_{\rm CDM}h^2$. So, at $95\%$ c.l., we take
\beq \Omega_{\rm LSP}h^2\lesssim0.12.\label{cdmb}\eeq
An upper bound on $m_{\rm LSP}$ (or $m_{\tilde\chi}$) can be
derived from Eq.~(\ref{cdmb}) since, in general, $\Omx$ increases
with $\mx$. The calculation of $\Omx$ in \mcr\ includes accurately
thermally averaged exact tree-level cross sections of all the
possible (co)annihilation processes \cite{cmssm1, cdm}, treats
poles \cite{cmssm2, qcdm, nra} properly, and uses one-loop QCD and
SUSY QCD corrections \cite{copw, qcdm, pallis, microbsg} to the
Higgs decay widths and couplings to fermions.

\paragraph{\hspace*{-0.3cm}
The Branching Ratio $\bsg$ of $b\to s\gamma$.}
The most recent experimental world average for
${\rm BR}(b\rightarrow s\gamma)$ is known \cite{bsgexp} to be
$\lf3.52\pm0.23\pm0.09\rg\times10^{-4}$ and its updated SM
prediction is $\lf3.15\pm0.23\rg\times10^{-4}$ \cite{bsgSM}.
Combining in quadrature the experimental and theoretical errors
involved, we obtain the following constraints on this
branching ratio at $95\%$ c.l.:
\beq 2.84\times 10^{-4}\lesssim \bsg \lesssim 4.2\times 10^{-4}.
\label{bsgb} \eeq
The computation of $\bsg$ in the {\tt micrOMEGAs} package
presented in \cref{microbsg} includes \cite{nlobsg}
\emph{next-to-leading order} (NLO) QCD corrections to the charged
Higgs boson ($H^\pm$) contribution, the $\tan\beta$ enhanced
contributions, and resummed NLO SUSY QCD corrections. The $H^\pm$
contribution interferes constructively with the SM contribution,
whereas the SUSY contribution interferes destructively with the
other two contributions for $\mu>0$. The SM plus the $H^\pm$ and
SUSY contributions initially increases with $\mx$ and yields a
lower bound on $\mx$ from the lower bound in Eq.~(\ref{bsgb}).
(For higher values of $\mx$, it starts mildly decreasing.)

\paragraph{\hspace*{-0.3cm} The Branching Ratio $\bmm$ of
$B_s\to\mu^+\mu^-$.}
The rare decay $B_s\to \mu^+\mu^-$ occurs via $Z$ penguin and
box diagrams in the SM and, thus, its branching ratio is
highly suppressed. The SUSY contribution, though, originating
\cite{bsmm, mahmoudi} from neutral Higgs bosons in chargino-,
$H^\pm$-, and $W^\pm$-mediated penguins behaves as
$\tan^6\beta/m^4_A$ and hence is particularly important for large
$\tan\beta$'s. We impose the following $95\%$ c.l. upper bound:
\beq \bmm\lesssim5.8\times10^{-8} \label{bmmb} \eeq
as reported \cite{bmmexp} by the CDF collaboration. This bound
implies a lower bound on $\mx$ since $\bmm$ decreases as
$m_{\rm LSP}$ increases.

\paragraph{\hspace*{-0.3cm} The Branching Ratio
${\rm BR}\lf B_u\to \tau\nu\rg$ of $B_u\to \tau\nu$.}
The purely leptonic decay
$B_u\to \tau\nu$ proceeds via $W^\pm$- and $H^\pm$-mediated
annihilation processes. The SUSY contribution, contrary to the
SM one, is not helicity suppressed and depends on the mass
$m_{H^\pm}$ of the charged Higgs boson since it behaves
\cite{Btn, mahmoudi} as $\tan^4\beta/m^4_{H^\pm}$. The ratio
$\btn$ of the CMSSM to the SM branching ratio of
$B_u\to \tau\nu$ increases with $\mx$ and approaches unity. It
is to be consistent with the following $95\%$ c.l. range
\cite{bsgexp}:
\beq 0.52\lesssim\btn\lesssim2.04\ .\label{btnb} \eeq
A lower bound on $\mx$ can be derived from the lower bound in
this inequality.

\paragraph{\hspace*{-0.3cm} Muon Anomalous Magnetic Moment.}
\label{phenoc}
The quantity $\delta a_\mu$, which is defined in \Sref{sec:intro},
can be attributed to SUSY contributions arising from
chargino-sneutrino and neutralino-smuon loops. The relevant
calculation is based on the formulas of Ref.~\cite{gmuon}. The
absolute value of the result decreases as $m_{\rm LSP}$ increases
and its sign is positive for $\mu>0$. On the other hand, the
calculation of $a^{\rm SM}_\mu$ is not yet stabilized mainly
because of the ambiguities in the calculation of the hadronic
vacuum-polarization contribution. According to the  most
up-to-date evaluation of this contribution in Ref.~\cite{g2davier},
there is still a discrepancy between the findings based on the
$e^+e^-$-annihilation data and the ones based on the
$\tau$-decay data. Taking into account the more reliable calculation
based on the $e^+e^-$ data and the experimental measurements
\cite{g2exp} of $a_\mu$, we obtain the following $95\%$ c.l. range:
\beq~12.7\times 10^{-10}\lesssim \delta a_\mu\lesssim 44.7\times 10^{-10}.
\label{g2b}\eeq
The $\tau$-decay based calculation, on the other hand, yields
the following $95\%$ c.l. range:
\beq~2.9\times 10^{-10}\lesssim \delta a_\mu\lesssim 36.1\times 10^{-10}.
\label{g2btau}\eeq
A lower [upper] bound on $\mx$ can be derived from the upper [lower]
bound in \Erefs{g2b}{g2btau}. As it turns out, only the upper bound
on $\mx$ is relevant in our case. Taking into account the
aforementioned computational instabilities, we will impose the less
stringent upper bound on $\mx$ from the $\tau$-decay based calculation.
However, we will also depict the more stringent bound from the
$e^+e^-$-annihilation data for comparison.

\paragraph{Collider Bounds.} For our analysis, the only relevant
collider bound is the $95\%$ c.l. LEP bound \cite{lepmh} on the
lightest CP-even neutral Higgs boson mass
\beq m_h\gtrsim114.4~{\rm GeV},\label{mhb} \eeq
which gives a lower bound on $m_{\rm LSP}$. The calculation of
$m_h$ in the package {\tt SOFTSUSY} \cite{Softsusy} includes the
full one-loop SUSY corrections and some zero-momentum two-loop
corrections \cite{2loops}. The results are well tested
\cite{comparisons} against other spectrum calculators.

\begin{figure}[!t]
\includegraphics[width=65mm,angle=-90]{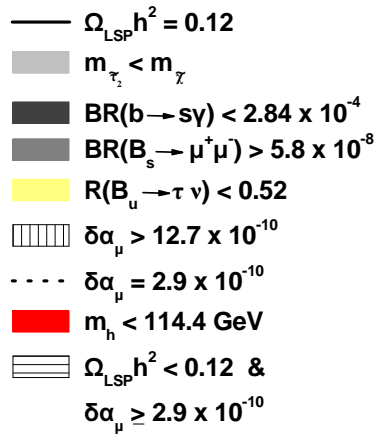}
\caption{Summary of the conventions adopted in Figs.~\ref{Mmx} and
\ref{AMgx} for the various restrictions on the parameters of the
model.}
\label{Capgx}
\end{figure}

\begin{figure*}[!tb]
\centering
\includegraphics[width=65mm,angle=-90]{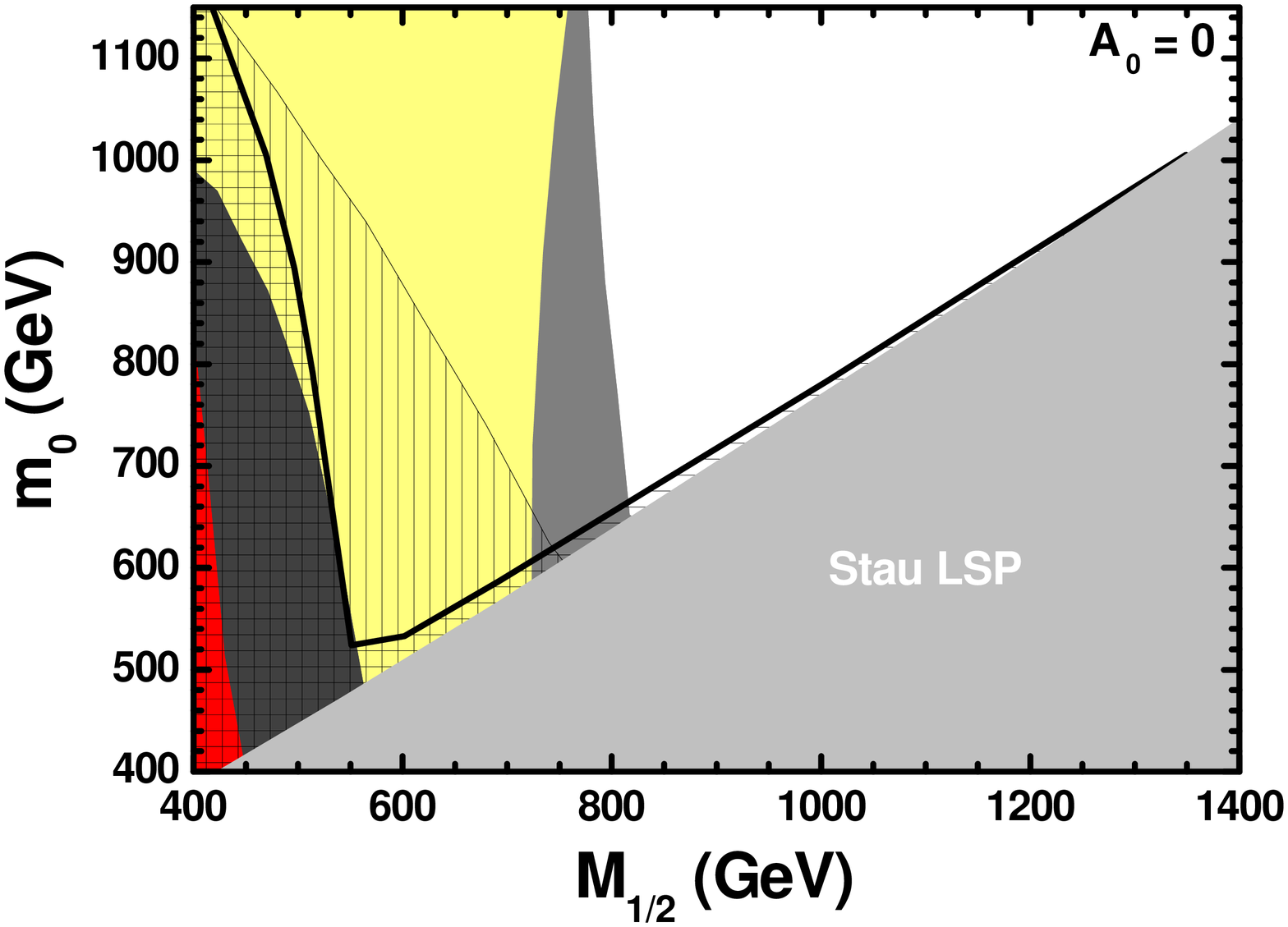}
\includegraphics[width=65mm,angle=-90]{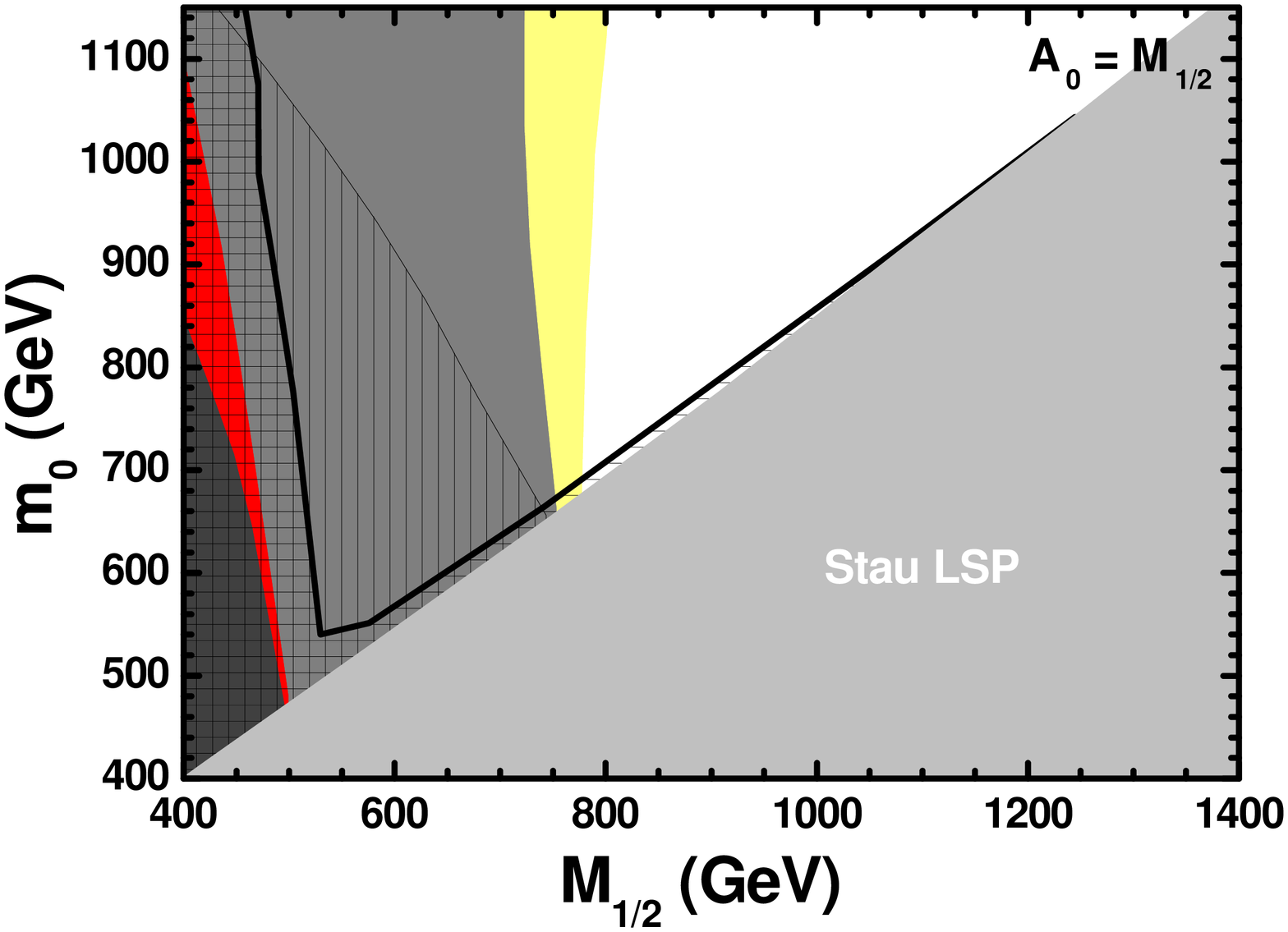}\\
\includegraphics[width=65mm,angle=-90]{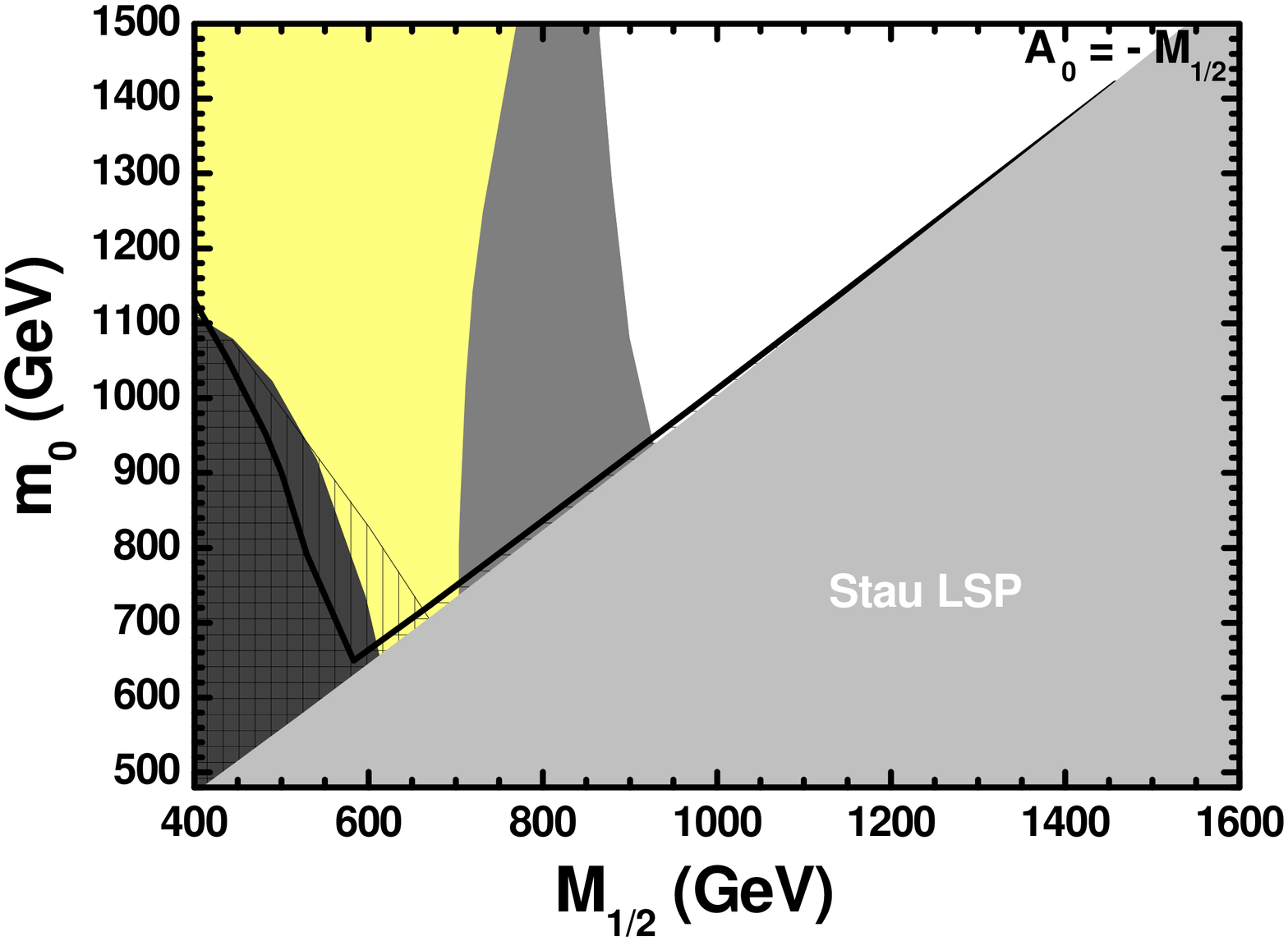}
\includegraphics[width=65mm,angle=-90]{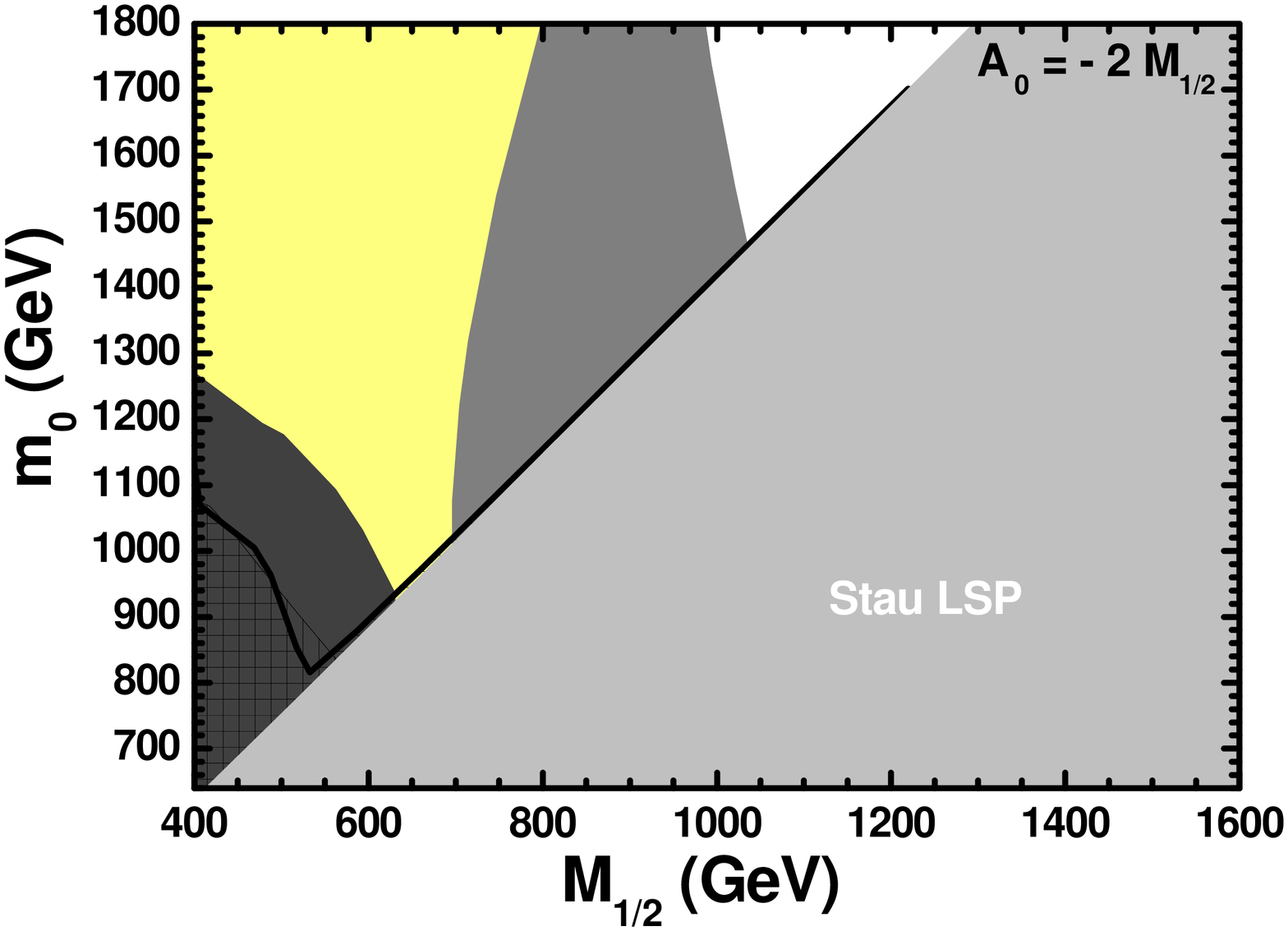}
\caption{Restrictions in the $\Mg-m_{0}$ plane for
various values of $A_0/\Mg$ indicated in the graphs. The
conventions adopted are described in \Fref{Capgx}.}
\label{Mmx}
\end{figure*}

\section{Restrictions on the SUSY Parameters} \label{results}

\begin{figure}[!t]
\includegraphics[width=65mm,angle=-90]{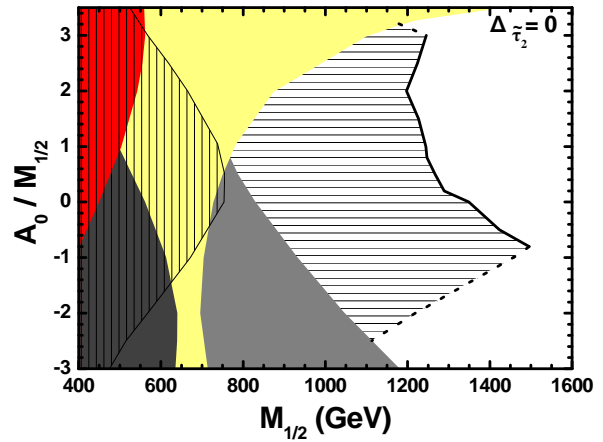}
\caption{Restrictions in the
$M_{1/2}-A_{0}/M_{1/2}$ plane for $\Delta_{\tilde\tau_2}=0$
following the conventions of \Fref{Capgx}, but with the
horizontally hatched region not extended to areas excluded by
other constraints.}
\label{AMgx}
\end{figure}

Imposing the requirements described above, we can delineate the
allowed parameter space of our model. The predicted mass spectra
are possibly relevant for the LHC searches. Throughout our
investigation, we consider the central values for the SM
parameters $M_t$, $m_b(M_Z)$, $m_\tau(M_Z)$, and $\alpha_s(M_Z)$.
We adopt the following conventions for the various lines and
regions in the relevant figures (Figs.~\ref{Mmx} and \ref{AMgx})
-- see \Fref{Capgx}:

\begin{itemize}

\item on the solid black line, \Eref{cdmb} is saturated,

\item the light gray region is cosmologically excluded
since it predicts charged LSP,

\item the dark gray region is excluded by the lower bound
in \Eref{bsgb},

\item the gray region is excluded by \Eref{bmmb},

\item the yellow region is excluded by the lower bound
in \Eref{btnb},

\item the vertically hatched region is favored by
the lower bound in \Eref{g2b},

\item on the dotted black line, the lower bound in
\Eref{g2btau} is saturated,

\item the red region is excluded by \Eref{mhb},

\item the horizontally hatched region is allowed
by both \Eref{cdmb} and the lower bound in
\Eref{g2btau}.

\end{itemize}
Note that the upper bounds in Eqs.~(\ref{bsgb}), (\ref{btnb}),
(\ref{g2b}), and (\ref{g2btau}) do not restrict the parameters of
our model.

We present the restrictions from all the requirements imposed
in the $\Mg-m_0$ plane for $A_0/\Mg=0$, 1, $-1$, and $-2$ in
\Fref{Mmx}. We remark that the lower bound on $\Mg$ comes from
\Eref{bmmb} for $A_0/\Mg=0$, $-1$, and $-2$ and from the lower
bound in \Eref{btnb} for $A_0/\Mg=1$. Also, from the relevant
data, we observe that the lower bound in \Eref{btnb} is
fulfilled for the mass of the CP-odd Higgs boson
$m_A\simeq520~\GeV$ and almost independently of the other
parameters. Finally, note that, for $\AMg=-1$ and $-2$, the
bound in \Eref{mhb} is violated for $M_{1/2}<400~\GeV$ and,
consequently, does not appear in the relevant diagrams.

The constraint in \Eref{cdmb} is, in general, satisfied in
two well-defined distinct regions in the diagrams of
\Fref{Mmx}. In particular,

\begin{itemize}

\item the region to the left of the almost vertical part of
the line corresponding to the upper bound on $\Mg$ from
\Eref{cdmb}, where the LSP annihilation via the $s$-channel
exchange of a CP-odd Higgs boson $A$ is by far the dominant
(co)annihilation process. However, this region is excluded by
the constraints in Eqs.~(\ref{bmmb}) and (\ref{btnb}). On the
other hand, it is well known -- see e.g. Refs.~\cite{cmssm2, qcdm}
-- that this region is extremely sensitive to variations of
$m_b(M_Z)$. Indeed, we find that as $m_b(M_Z)$ decreases, the
$A$-boson mass $m_A$ increases and approaches $2m_{\rm LSP}$.
The $A$-pole neutralino annihilation is then enhanced and
$\Omega_{\rm LSP}h^2$ is drastically reduced causing an
increase of the upper bound on $\Mg$. However, even if we
reduce $m_b(M_Z)$, we do not find any $A$-pole annihilation
region which is allowed by the requirements of
Eqs.~(\ref{bmmb}) and (\ref{btnb}).

\item the narrow region which lies just above the light gray
area with charged LSP, where bino-stau coannihilations
\cite{cmssm1,cdm} take over
leading to a very pronounced reduction of $\Omx$. A large
portion of this region survives after the application of the
requirements in Eqs.~(\ref{bmmb}) and (\ref{btnb}) and
constitutes the overall allowed parameter range of our model
for the given $A_0$. To get a better understanding of this area,
we can replace the parameter $m_0$ by the relative mass
splitting $\Dst$ between the LSP and the lightest stau, defined
in Sec.~\ref{sec:intro}. We observe that the overall allowed
region requires $\Dst\lesssim0.025$. It is evident from
\Fref{Mmx} that the slope of the boundary line with $\Dst=0$
increases as $\AMg$ moves away from zero in both directions.
Note that this slope in our model turns out to be larger than
the one obtained in other versions of the CMSSM --
cf. \cref{cmssm1} -- with lower values of $\tan\beta$. As a
consequence, small variations of $m_0$ or $\Mg$ lead, in our
model, to more drastic variations in $\Dst$.

\end{itemize}
Finally, we note that the more stringent upper bound on $\Mg$ from
the lower bound in \Eref{g2b} is not satisfied for the values
taken for $\AMg$ in \Fref{Mmx}, with the values $\AMg=0$ and $1$
being much more favored. On the other hand, the lower bound in
\Eref{g2btau} is fulfilled in the whole allowed region for
$\AMg=0$ and $1$ whereas, for $\AMg=-1$ and $-2$, it imposes an
upper bound on $\Mg$ which overshadows the bound on $\Mg$ from
\Eref{cdmb}. Since the saturation of the lower bound in \Eref{g2btau}
occurs for $\Dst\lesssim 0.01$, the portion of the dotted black
line -- see Fig.~\ref{Capgx} -- which connects the black solid
line with the boundary of the gray area is not visible in the
relevant panels of \Fref{Mmx}.

To get a better idea of the allowed parameter space, we focus
on the coannihilation regime and construct the allowed region
in the $\Mg-A_0/\Mg$ plane. This is shown in \Fref{AMgx},
where we depict the restrictions on the parameters from the
various constraints for $\Dst=0$. This choice ensures the
maximal possible reduction of $\Omx$ due to the
$\tilde\chi-\tilde\tau_2$ coannihilation. So, for $\Dst=0$, we
find the maximal $\Mg$ or $\mx$ allowed by \Eref{cdmb} for a
given value of $A_0/\Mg$. We observe that, for $-0.8\lesssim
A_0/\Mg\lesssim3$ [$-2.55\lesssim A_0/\Mg\lesssim-0.8$ and
$3\lesssim A_0/\Mg\lesssim3.21$] the overall upper bound on
$\Mg$ or $\mx$ is derived from the bound in Eq.~(\ref{cdmb})
[lower bound in \Eref{g2btau}]. We find that, for $A_0/\Mg<0$,
processes with $\tilde\tau_2\tilde\tau_2^\ast$ in the initial
state and $W^\pm W^\mp$, $W^\pm H^\mp$ in the final one become
more efficient (with a total contribution to the effective cross
section of about $14$ to $21\%$ as $A_0/\Mg$ decreases from 0 to
-2.55) and so coannihilation is strengthened and $m_{\rm LSP}$'s
larger than in the $A_0/\Mg>0$ case are allowed. The overall
maximal $\Mg\simeq1495.4~\GeV$ or $\mx\simeq677~\GeV$ is
encountered at $A_0/\Mg\simeq-0.8$. On the other hand, for
$-2.55\lesssim A_0/\Mg\lesssim0.7$ [$0.7\lesssim A_0/\Mg
\lesssim3.21$] the lower bound on $\Mg$ or $\mx$ is derived
from the bound in Eq.~(\ref{bmmb}) [lower bound in \Eref{btnb}].
The overall allowed lowest $\Mg\simeq771.22~\GeV$ or
$\mx\simeq341~\GeV$ is encountered at $A_0/\Mg\simeq0.7$.
Let us remark that the more stringent upper bound on
$\Mg$ from the lower bound in \Eref{g2b} is not satisfied in
the allowed region of our model, since there is no common
region between the horizontally and the vertically hatched areas
for any $A_0/\Mg$. However, for $0\lesssim A_0/\Mg\lesssim1$,
these areas are quite close to each other. Note that
increasing $\Dst$ within its
allowed range $0-0.025$ does not alter the boundaries of the
various constraints in any essential way, except the solid line
which is displaced to the left so that the allowed area shrinks
considerably.

The deviation from YU can be estimated by defining \cite{nova}
the relative splittings $\delta h_{b}$ and $\delta h_{\tau}$ at
$M_{\rm GUT}$ through the relations:
\beq \delta h_{b}\equiv\frac{h_{b}-h_t}{h_t}=-\frac{2c}{1+c}=
-\delta h_{\tau}\equiv\frac{h_t-h_{\tau}}{h_t}\cdot\eeq
In the allowed (horizontally hatched) area of \Fref{AMgx}, the
ranges of the parameters $c,~\delta h_{\tau},~\delta h_{b}$, and
$\tan\beta$ are
\bea \nonumber 0.149\lesssim c\lesssim0.168,\\ \nonumber
0.26\lesssim\delta h_{\tau}=-\delta h_b \lesssim0.29,\\
56.3\lesssim \tan\beta\lesssim 57.7. \eea
Let us underline that, although the required deviation from YU
is not so small, the restrictions from YU are not completely lost
since $\tan\beta$ remains large -- close to 60 -- and the
deviation from YU is generated in a GUT-inspired well-motivated
way.

Taking into account the results depicted in \Fref{AMgx}, we can
make predictions for the sparticle and the Higgs boson spectrum
of our model, which may be observable at the LHC. In
\Tref{spectrum}, we list the model input and output parameters,
the masses in $\GeV$ of the sparticles -- neutralinos
$\tilde\chi$, $\tilde\chi_2^{0}$, $\tilde{\chi}_{3}^{0}$,
$\tilde{\chi}_{4}^{0}$, charginos $\tilde{\chi}_{1}^{\pm}$,
$\tilde{\chi}_{2}^{\pm}$, gluinos $\tilde{g}$, squarks
$\tilde{t}_1$, $\tilde{t}_2$, $\tilde{b}_1$, $\tilde{b}_2$,
$\tilde{u}_{L}$, $\tilde{u}_{R}$, $\tilde{d}_{L}$,
$\tilde{d}_{R}$, and sleptons $\tilde\tau_1$, $\tilde\tau_2$,
$\tilde\nu_\tau$, $\tilde{e}_L$, $\tilde{e}_R$, $\tilde{\nu}_{e}$
-- and the Higgs bosons ($h$, $H$, $H^\pm$, $A$), and the values
of the various low energy observables for $A_0/\Mg=0$, $\pm1$,
and $-2$ and for the lowest possible $\Mg$ in each case adjusting
$\Dst$ so as $\Omx\simeq 0.11$. Note that we consider the squarks
and sleptons of the two first generations as degenerate. From the
values of the various observable quantities, it is easy to
verify that all the relevant constraints are met. We also included
in \Tref{spectrum} predictions for the possible direct detection
of the LSP using central values for the hadronic inputs
$f_{{\rm T}q}^p$ or $\Delta_{q}^{p}$ -- see \Sref{det}.

\begin{table}[!t]
\caption{Input and output parameters, masses of the sparticles and
Higgs bosons, and values of the low energy observables of our
model for four values of $A_0/\Mg$. Recall that
$1~\pb\simeq2.6\times10^{-9}~\GeV^{-2}$.} \vspace*{1.mm}

\begin{tabular}{c@{\hspace{0.1cm}}c@{\hspace{0.3cm}}c@{\hspace{0.3cm}}c@{\hspace{0.3cm}}c}
\toprule \multicolumn{5}{c}{Input parameters}\\\colrule
$A_0/\Mg$ &$0$ &$1$&$-1$ &$-2$ \\
$c$ & $0.161$ &$156$&$0.165$ &$0.168$\\
$\Mg/\GeV$ & $825.7$ &$776.06$&$927.25$ &$1041.8$ \\
$m_0/\GeV$ &$665.4$ &$687.5$&$943.1$&$1466.8$ \\ \colrule
\multicolumn{5}{c}{Output parameters}\\\colrule
$\tan\beta$ & $57$ & $56.8$ & $57.4$ & $57.7$\\
$h_t(M_{\rm GUT})$ & $0.58$ & $0.58$ & $0.58$ & $0.58$\\
$100\delta h_\tau(M_{\rm GUT})$ & $27.7$ & $26.9$ & $28.3$ & $28.7$\\
$\mu/\GeV$ & $925.8$ & $804$ & $1170$ & $1505$\\
$\Dst (\%)$ & $2.46$ & $2.45$ & $2.13$ & $1.52$\\
\colrule
\multicolumn{5}{c}{Masses in ${\rm GeV}$ of sparticles and
Higgs bosons}\\\colrule
$\tilde\chi$& $365.7$ &$342.7$ &$413.2$ &$467.5$\\
$\tilde\chi_2^{0}$ &$705$&$656$ &$802$ &$909$\\
$\tilde{\chi}_{3}^{0}$ &$927$&$807$ &$1170$ &$1502$\\
$\tilde{\chi}_{4}^{0}$ &$940$ &$827$ &$1177$ &$1506$\\
$\tilde{\chi}_{1}^{\pm}$ &$940$ &$827$&$1177$ &$1506$\\
$\tilde{\chi}_{2}^{\pm}$ & $705$ &$656$ &$802$ &$909$ \\
$\tilde{g}$&$1916$&$1813$&$2145$&$2412$ \\ \colrule
%
$\tilde{t}_1$&$1585$&$1530$&$1752$&$1980$ \\
$\tilde{t}_2$  &$1383$ &$1352$ &$1506$&$1666$\\
$\tilde{b}_1$ &$1578$ &$1526$ &$1752$&$2008$\\
$\tilde{b}_2$ &$1498$ &$1454$&$1670$&$1916$ \\
$\tilde{u}_{L}$&$1052$&$1762$ &$2134$&$2585$ \\
$\tilde{u}_{R}$ &$1011$ &$1694$ &$2054$&$2503$ \\
$\tilde{d}_{L}$& $1055$ &$1764$ &$2135$ & $2586$\\
$\tilde{d}_{R}$&$1006$&$1764$&$2045$&$2494$\\\colrule
$\tilde\tau_1$&$777$&$754$&$956$&$1283$\\
$\tilde\tau_2$& $374.7$&$351.1$&$422$&$474.6$\\
$\tilde\nu_\tau$ &$756$&$738$&$939$&$1272$\\
$\tilde{e}_L$&$880$&$875$&$1142$&$1635$\\
$\tilde{e}_R$&$740$&$752$&$1010$&$1523$\\
$\tilde{\nu}_{e}$ &$876$ &$871$ &$1139$&$1633$\\\colrule
%
$h$   &$118.1$ &$117$ &$119.7$&$121.3$\\
$H$ &$584$  &$519$ &$668$&$747.6$\\
$H^{\pm}$   &$591$  &$527$ &$674$&$752.8$ \\
$A$  &$585$ & $520$&$669$ &$748$\\\colrule
\multicolumn{5}{c}{Low energy observables}\\\colrule
%
$10^4\bsg$ &$3.32$  &$3.41$ &$3.32$&$3.37$\\
$10^8\bmm$   &$5.76$ &$5.3$ &$5.78$&$5.8$\\
$\btn$   &$0.61$  &$0.52$ &$0.69$&$0.74$\\
$10^{10}\Dam$  &$10.6$&$11.6$ &$6.9$ &$3.9$\\\colrule
$\Omx$ &$0.11$&$0.11$&$0.11$&$0.11$\\[1mm]
$\ssi / 10^{-9} \pb$ &$0.536$&$1.1$ &$0.2$&$0.076$\\[1mm]
$\ssd /  10^{-7} \pb$ &$1.96$&$4.1$ &$0.6$&$0.2$\\[1mm]
\botrule
\end{tabular}
\label{spectrum}
\end{table}

\begin{figure}[t!]
\includegraphics[width=65mm,angle=-90]{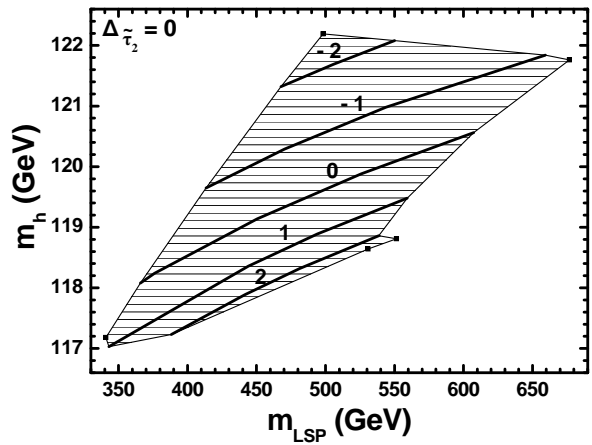}
\caption{The allowed (horizontally hatched) region
in the $m_{\rm LSP}-m_h$ plane for $\Dst\simeq0$. We also depict the
curves corresponding to various values of $A_0/M_{1/2}$, indicated
on them. The dark points on the boundary correspond to
$A_0/M_{1/2}=-2.55$, -0.8, 3, 3.21, and 0.8 starting from the point
at the top of the allowed area and moving clockwise.}
\label{mLSPhx}
\end{figure}

For the lowest masses of the Higgs and sparticle spectrum ($m_h$
and $\mx$), we present even more explicit predictions in
\Fref{mLSPhx}, where we depict the allowed $m_h$'s versus $\mx$
for $\Dst\simeq0$ and $A_0/\Mg=0$, $\pm1$, and $\pm2$. As can be
seen from \Fref{AMgx},
the lower limits on the solid lines for $A_0/\Mg=0$, $-1$, and $-2$
[$A_0/\Mg=1$ and $2$] are found from the bound in \Eref{bmmb}
[lower bound in \Eref{btnb}] -- see also Table I. On the other
hand, the upper limits of the solid lines for $\AMg=0$, $1$,
and $2$ [$\AMg=-1$ and $-2$] are found from the bound in
\Eref{cdmb} [lower bound in \Eref{g2btau}]. The approximate
overall allowed area in the $\mx-m_h$ plane is hatched. Shown
are also the boundary points of this region at $\AMg\simeq-2.55$,
$-0.8$, $3$, $3.21$, and $0.7$ starting from the point at the
top of the allowed area and moving clockwise.

As one can see from \Fref{mLSPhx}, $m_h$ increases with $\mx$
and as $A_0$ decreases. Since the maximum allowed $\mx$ from
the bound in \Eref{cdmb} or the lower bound in \Eref{g2btau} is
achieved at $\Dst\simeq0$ for given $A_0$, we conclude that the
maximum
possible allowed $m_h$ can be obtained for $\Dst\simeq0$. On
the other hand, the minimum possible allowed $m_h$ practically
coincides with its value for $\Dst\simeq0$ since variation of
$\Dst$ within the values allowed by \Eref{cdmb} causes minor
modifications of $m_h$ for fixed $\Mg$. For $A_0=0$, we find
$826.4\lesssim\Mg/\GeV\lesssim1348.9$ or
$365.9\lesssim\mx/\GeV\lesssim607.4$ and $118.1\lesssim m_h/
\GeV\lesssim120.6$. The overall minimum [maximum] $m_h$ is
$117.03$ [$122.2$] obtained at
$A_0/\Mg\simeq1$ [$A_0/\Mg\simeq-2.55$] for
$\Mg=776.6~\GeV$ [$\Mg=1106.6~\GeV$] or $\mx\simeq343.1~\GeV$
[$\mx\simeq498.3~\GeV$].

\begin{figure*}[!tb]
\centering
\includegraphics[width=65mm,angle=-90]{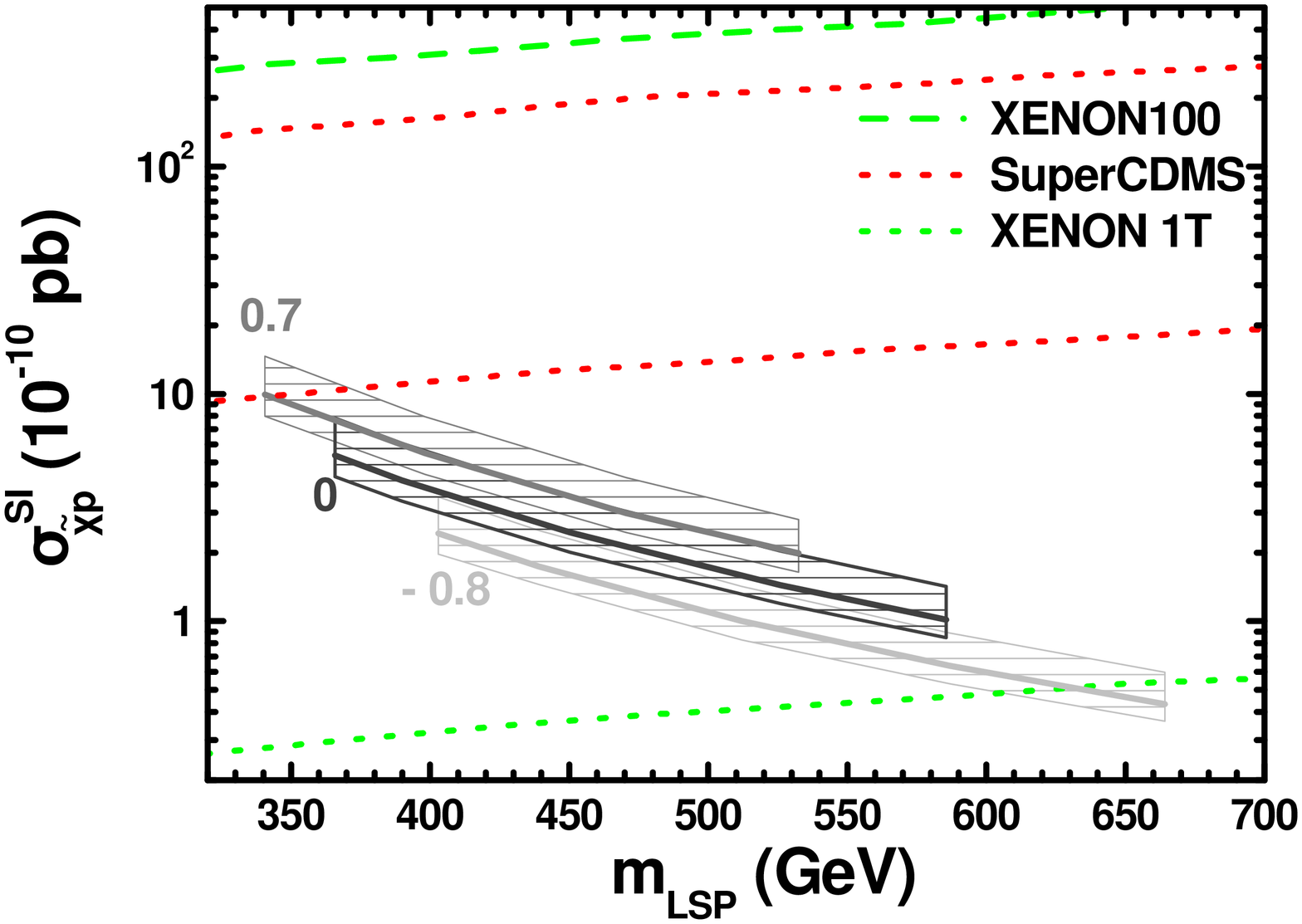}
\includegraphics[width=65mm,angle=-90]{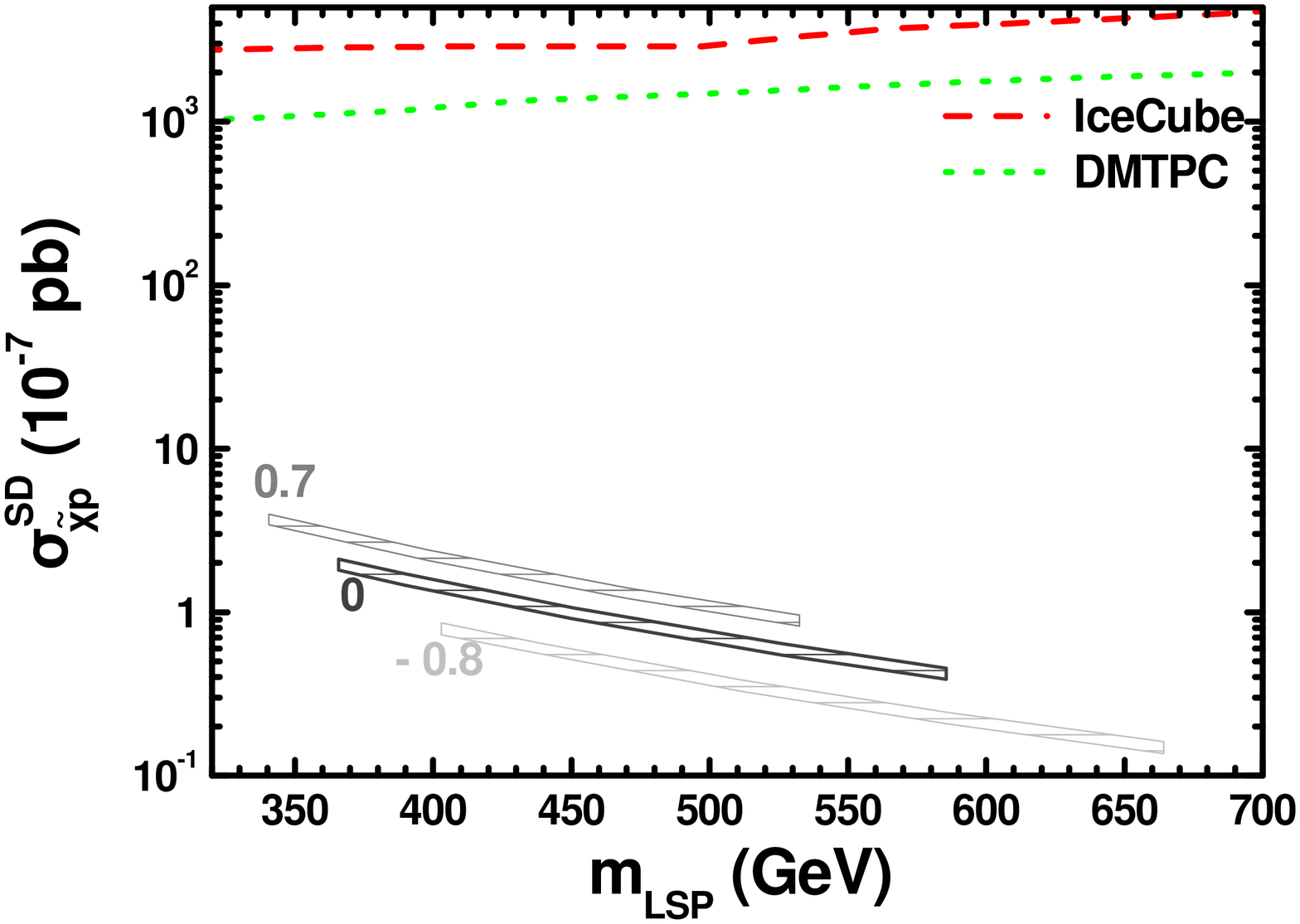}
\caption{The SI and SD $\tilde\chi-p$ cross sections
$\ssi$ and $\ssd$, respectively, versus $m_{\rm LSP}$ for various
$\AMg$'s indicated in the graphs. The bold solid lines in the
left panel are derived by fixing $\Omega_{\rm LSP}h^2$ and
$f_{{{\rm T}q}}^{p}$ to their central values
in Eqs.~(\ref{cdmba}) and (\ref{rgis4})-(\ref{rgis6}), whereas
the hatched bands in both panels by allowing the hadronic inputs
$f_{{{\rm T}q}}^{p}$ or  $\Delta_{q}^{p}$ to vary in their ranges
in Eqs.~(\ref{rgis4})-(\ref{rgis6}) or
(\ref{Dqp1})-(\ref{Dqp3}). The
present and planned sensitivity limits of the various experimental
projects are also depicted by dashed and dotted lines,
respectively.}
\label{detx}
\end{figure*}

\section{CDM Direct Detection} \label{det}

As we have shown, our model possesses a limited and well-defined
range of parameters allowed by all the relevant cosmological and
phenomenological constraints. It would be, thus, interesting to
investigate whether the predicted LSPs in the universe could be
detected in the current or planned direct CDM searches
\cite{Cdms2,Xenon,icecube}, which look for evidence of
weakly-interacting massive particles through scattering on nuclei.
The quantities which are conventionally used in the recent
literature for comparing experimental results and theoretical
predictions are the spin-independent (SI) and spin-dependent (SD)
lightest neutralino-proton ($\tilde\chi-p$) scattering cross
sections $\ssi$ and $\ssd$, respectively.

These quantities are calculated by employing the relevant routine
of the {\tt micrOMEGAs} package \cite{Detmicro} based on the full
one-loop treatment of Ref.~\cite{drees}, which happens to agree
with the tree-level approximation \cite{Detellis} for the values
of the SUSY parameters encountered in our model. Following the
approach of Refs.~\cite{Detmicro,Detellis}, we calculate the
scalar form factors for light quarks in the proton
$f^p_{{{\rm T}_q}}$ (with $q=u,d,s$), needed for the calculation
of $\ssi$, via the formulas:
\beqs\bea && \label{rgis1} f_{{\rm T}d}^p= \frac{2\spn}{m_p\lf1 + \frac{m_u}
{m_d} \rg\lf1 + \frac{B_u}{B_d}\rg},\\
   && \label{rgis2} f_{{\rm T}u}^p=\frac{m_u}{m_d}\frac{B_u}{B_d}f_{{\rm
T}_d}^p,\\
  && f_{{\rm T}s}^p=\frac{y\spn}{m_p\lf1 + \frac{m_u}{m_d}\rg}
  \frac{m_s}{m_d}.\label{rgis3}\eea\eeqs
Here we take for the mass of the proton $m_p=0.939~\GeV$ and for
the light quark mass ratios
\beq \frac{m_u}{m_d}=0.553\pm0.043~~\mbox{and}~~\frac{m_s}{
m_d}=18.9\pm0.8,
\eeq
whereas the ratio $B_u/B_d$ is evaluated from
\beq \frac{B_u}{B_d}=\frac{2z-(z-1)y}{2+(z-1)y}\eeq
with $z=1.49$. The uncertainties in $z$ and the quark mass ratios
are negligible compared to the uncertainties in the pion-nucleon
sigma term $\spn$ and the fractional strange quark content of the
nucleon $y$, for which recent lattice simulations suggest
\cite{lattice} that, at $68\%$ c.l.,
\beq \spn = 53^{+21.1}_{-7.3}~{\rm MeV} ~~\mbox{and}~~ y =
0.030^{+0.017}_{-0.018}.\eeq
Taking into account the relations above, we find the
following 1$\sigma$ ranges for the $f_{{\rm T}q}^p$'s:
\beqs\bea && \label{rgis4} f_{{\rm T}u}^p=0.024^{+0.0095}_{-0.0032},\\
   && \label{rgis5} f_{{\rm T}d}^p=0.029_{-0.0042}^{+0.012},\\
  && f_{{\rm T}s}^p=0.021^{+0.025}_{-0.013}. \label{rgis6}\eea\eeqs
Note that $f_{{\rm T}s}^p$ turns out to be considerably smaller
than its older value -- cf. \cref{nova} -- reducing thereby the
extracted $\ssi$.

For the calculation of $\ssd$, the relevant
axial-vector form factors for light quarks in the proton
$\Delta_q^{p}$ (with $q=u,d,s$) are taken to lie in their
1$\sigma$ ranges \cite{Dqpc}:
\beqs\begin{eqnarray}
&& \label{Dqp1} \Delta_{u}^{p}=+0.842\pm0.012,\\
&& \label{Dqp2} \Delta_{d}^{p}=-0.427\pm0.013, \\ &&
\Delta_{s}^{p}=-0.085\pm0.018.\label{Dqp3}
\end{eqnarray}\eeqs

Taking the central value of $\Omx$ in Eq.~(\ref{cdmba}), but
allowing the hadronic inputs $f_{{\rm T}q}^{p}$ or $\Delta_{q}^{p}$
to vary within their ranges in Eqs.~(\ref{rgis4})-(\ref{rgis6})
or (\ref{Dqp1})-(\ref{Dqp3}),
respectively, we derive the dark gray, gray, and light gray hatched
bands in the $\mx-\ssi$ or $\mx-\ssd$ plane corresponding to
$\AMg=0$, $0.7$, and $-0.8$, respectively -- Fig.~\ref{detx}. The
selected values of $\AMg$ allow us to cover the whole range of the
allowed $\mx$'s in our model -- cf. \Fref{mLSPhx}. The bold solid
lines in the middle of the bands of the left panel in Fig.~\ref{detx}
correspond to the central values of the $f_{{\rm T}q}^{p}$'s.
We used the central value of $\Omx$ since, as it turns out, for fixed
$\mx$, $\ssi$ and $\ssd$ are almost insensitive to the variation of
$\Omx$ within the range of Eq.~(\ref{cdmba}) -- or,  equivalently, to
the required variation of $\Dst$. The width of the bands is almost
exclusively due to the variation of $f_{{\rm T}q}^{p}$ or
$\Delta_{q}^{p}$. As a consequence, the bands in the $\mx-\ssi$
plane are wider than those in the $\mx-\ssd$ plane due to the larger
uncertainties involved in the determination of the
$f_{{\rm T}q}^{p}$'s. In the left panel of \Fref{detx}, we
depict by a dashed green line the recently announced \cite{Xenon}
upper bound on $\ssi$ from XENON which is slightly lower than the
one from CDMSII \cite{Cdms2}, which is not included in the panel.
We also draw with dotted red lines the projected
sensitivities of SuperCDMS at Soudan and SNOLAB \cite{dmtools}
-- from top to bottom. Our model can be ultimately tested by
XENON-1 ton, whose planned sensitivity \cite{dmtools}, depicted by
a green dotted line, covers almost the whole available parameter
space of the model. On the contrary, as can be easily deduced from
the right panel of \Fref{detx}, $\ssd$ in our model lies well below
the sensitivity of IceCube \cite{icecube} (assuming neutralino
annihilation into
$W^+W^-$) -- depicted by a dashed red line -- and the expected
limit from the large DMTPC detector \cite{dmtools}, denoted by a
dotted green line. Therefore, the LSPs predicted by our model can
be detectable in the future projects which will release data on
$\ssi$. Furthermore, the overall upper bound on $m_{\rm LSP}$
found in \Sref{results} -- $\mx\lesssim677~\GeV$ -- implies
lower bounds on $\ssi$ and $\ssd$. Namely,
\beq \ssi\gtrsim
4.3~(3.6)\times10^{-11}~\pb~~\mbox{and}~~\ssd\gtrsim
1.5~(1.4)\times10^{-8}~\pb,
\label{sgmxp}\eeq
where the bounds in parentheses are derived by allowing
the $f_{{\rm T}q}^{p}$'s and $\Delta_q^p$'s to vary within
1$\sigma$. Needless to say that the low values of $\ssi$ and
$\ssd$ obtained here are due to the fact that we use universal
`asymptotic' gaugino  masses and, thus, the LSP is an almost
pure bino, as in every version of the CMSSM.

\section{Conclusions} \label{con}

We performed a revised scan of the parameter space of the
CMSSM with $\mu>0$ applying a suitable Yukawa
quasi-unification condition predicted by the SUSY GUT model
of Ref.~\cite{qcdm}, which has been constructed in order
to remedy the $b$-quark mass problem arising from exact
Yukawa unification and universal boundary conditions. We
took into account updated constraints from collider and
cosmological data. These constraints originate from the
CDM abundance in the universe, $B$ physics
($b \rightarrow s\gamma$, $B_s\to \mu^+\mu^-$, and
$B_u\to\tau\nu$), $\delta\alpha_\mu$, and $m_h$. We showed
that our model possesses a limited but not unnaturally
small range of parameters which is consistent with all
these requirements. Namely, the constraint arising from CDM
considerations can be satisfied simultaneously with all the
other constraints thanks to the drastic reduction of the
LSP relic density by neutralino-stau coannihilations.
For $A_0=0$, we find $365.9\lesssim\mx/\GeV\lesssim607.4$
and $118.1\lesssim m_h/{\rm GeV}\lesssim120.6$, whereas, in the
overall allowed region of our  model, we have
$-2.55\lesssim A_0/\Mg\lesssim3.21$ with
$341\lesssim\mx/\GeV\lesssim677$ and
$117\lesssim m_h/\GeV\lesssim 122.2$. Almost all the allowed
parameter space of our model will be accessible in future CDM
direct experiments which look for SI cross sections between
neutralino and proton.

It is worth mentioning that the present investigation constitutes
an improved version of the analysis in Ref.~\cite{qcdm}. The
consideration of the constraints from $\bmm$ and $\btn$, the
updated experimental results for all the other constraints, and
the evaluation of the particle spectrum employing {\tt SOFTSUSY}
are the main improvements in this work. The results obtained are
significantly different from the previous ones.

\section*{Note Added}  While this work was under completion, we
became aware of \cref{Shafi11}, where the CMSSM with Yukawa
quasi-unification is also analyzed. Although our results as
regards $\tan\beta$, $c$, $\ssi$, and $\ssd$ are similar, there
are large discrepancies as regards the CMSSM mass parameters and,
consequently, the mass spectrum. In particular, the ratio
$(m_A-2\mx)/2\mx$, which determines the strength of the $A$-pole
effect in reducing $\Omx$, is not allowed to be lower than $0.2$
in our case and, thus, this effect is excluded. Indeed, the
portion of the parameter space allowed by \Eref{cdmb} due to
$A$-pole neutralino annihilations is excluded by the $B$-physics
constraints in our analysis -- contrary to the findings of
\cref{Shafi11}. These discrepancies can be possibly attributed to
the fact that we use different numerical routines for the
calculation of both the SUSY spectra and the low energy
observables. It is well known \cite{comparisons2} that the
predictions of the various SUSY spectrum calculators do not
coincide in the large $\tan\beta$ regime. Our results as regards
the implementation of the electroweak symmetry breaking are
consistent with our initial investigation in \cref{qcdm}.

\acknowledgments We would like to thank G.~B\'{e}langer,
M.E.~G\'omez, S.~Heinemeyer, A.~Pukhov, and P.~Slavich for
enlightening correspondence as well as I.~Gogoladze and Q.~Shafi
for useful discussions related to \cref{Shafi11}. This work was
supported by the European Union under the Marie Curie Initial
Training Network `UNILHC' PITN-GA-2009-237920 and also by the
European Union (European Social Fund - ESF) and Greek national
funds through the Operational Program ``Education and Lifelong
Learning'' of the National Strategic Reference Framework (NSRF) -
Research Funding Program: Heracleitus II. Investing in knowledge
society through the European Social Fund.

\def\ijmp#1#2#3{{Int. Jour. Mod. Phys.}
{\bf #1},~#3~(#2)}
\def\plb#1#2#3{{Phys. Lett. B }{\bf #1},~#3~(#2)}
\def\zpc#1#2#3{{Z. Phys. C }{\bf #1},~#3~(#2)}
\def\prl#1#2#3{{Phys. Rev. Lett.}
{\bf #1},~#3~(#2)}
\def\rmp#1#2#3{{Rev. Mod. Phys.}
{\bf #1},~#3~(#2)}
\def\prep#1#2#3{{Phys. Rep. }{\bf #1},~#3~(#2)}
\def\prd#1#2#3{{Phys. Rev. D }{\bf #1},~#3~(#2)}
\def\npb#1#2#3{{Nucl. Phys. }{\bf B#1},~#3~(#2)}
\def\npps#1#2#3{{Nucl. Phys. B (Proc. Sup.)}
{\bf #1},~#3~(#2)}
\def\mpl#1#2#3{{Mod. Phys. Lett.}
{\bf #1},~#3~(#2)}
\def\arnps#1#2#3{{Annu. Rev. Nucl. Part. Sci.}
{\bf #1},~#3~(#2)}
\def\sjnp#1#2#3{{Sov. J. Nucl. Phys.}
{\bf #1},~#3~(#2)}
\def\jetp#1#2#3{{JETP Lett. }{\bf #1},~#3~(#2)}
\def\app#1#2#3{{Acta Phys. Polon.}
{\bf #1},~#3~(#2)}
\def\rnc#1#2#3{{Riv. Nuovo Cim.}
{\bf #1},~#3~(#2)}
\def\ap#1#2#3{{Ann. Phys. }{\bf #1},~#3~(#2)}
\def\ptp#1#2#3{{Prog. Theor. Phys.}
{\bf #1},~#3~(#2)}
\def\apjl#1#2#3{{Astrophys. J. Lett.}
{\bf #1},~#3~(#2)}
\def\apjs#1#2#3{{Astrophys. J. Suppl.}
{\bf #1},~#3~(#2)}
\def\n#1#2#3{{Nature }{\bf #1},~#3~(#2)}
\def\apj#1#2#3{{Astrophys. J.}
{\bf #1},~#3~(#2)}
\def\anj#1#2#3{{Astron. J. }{\bf #1},~#3~(#2)}
\def\mnras#1#2#3{{MNRAS }{\bf #1},~#3~(#2)}
\def\grg#1#2#3{{Gen. Rel. Grav.}
{\bf #1},~#3~(#2)}
\def\s#1#2#3{{Science }{\bf #1},~#3~(#2)}
\def\baas#1#2#3{{Bull. Am. Astron. Soc.}
{\bf #1},~#3~(#2)}
\def\ibid#1#2#3{{\it ibid. }{\bf #1},~#3~(#2)}
\def\cpc#1#2#3{{Comput. Phys. Commun.}
{\bf #1},~#3~(#2)}
\def\astp#1#2#3{{Astropart. Phys.}
{\bf #1},~#3~(#2)}
\def\epjc#1#2#3{{Eur. Phys. J. C}
{\bf #1},~#3~(#2)}
\def\nima#1#2#3{{Nucl. Instrum. Meth. A}
{\bf #1},~#3~(#2)}
\def\jhep#1#2#3{{J. High Energy Phys.}
{\bf #1},~#3~(#2)}
\def\jcap#1#2#3{{J. Cosmol. Astropart. Phys.}
{\bf #1},~#3~(#2)}

\end{document}